\documentclass{article}
\usepackage{ijcai22}

\usepackage{times}

\usepackage{soul}
\usepackage{url}
\usepackage[hidelinks]{hyperref}
\usepackage[utf8]{inputenc}
\usepackage[small]{caption}
\usepackage{graphicx}
\usepackage{bbding}
\usepackage{amsmath}
\usepackage{amssymb}
\usepackage{booktabs}
\usepackage{multirow}
\usepackage{algorithm}
\usepackage{algorithmic}
\usepackage{float}
\usepackage{pifont}
\newcommand{\cmark}{\ding{51}}%
\newcommand{\xmark}{\ding{55}}%

\title{A Novel Multi-Task Learning Method for Symbolic Music Emotion Recognition}

\author{
Jibao Qiu$^1$\and
C. L. Philip Chen$^1$\And
Tong Zhang$^{1}$ \footnote{Contact Author} \and
\affiliations
$^1$South China University of Technology\\
\emails
csj.b.qiu@mail.scut.edu.cn, 
\{philipchen, tony\}@scut.edu.cn
}

\begin{document}

\maketitle

\begin{abstract}
Symbolic Music Emotion Recognition(SMER) is to predict music emotion from symbolic data, such as MIDI and MusicXML. Previous work mainly focused on learning better representation via (mask) language model pre-training but ignored the intrinsic structure of the music, which is extremely important to the emotional expression of music. In this paper, we present a simple multi-task framework for SMER, which incorporates the emotion recognition task with other emotion-related auxiliary tasks derived from the intrinsic structure of the music. The results show that our multi-task framework can be adapted to different models. Moreover, the labels of auxiliary tasks are easy to be obtained, which means our multi-task methods do not require manually annotated labels other than emotion. Conducting on two publicly available datasets (EMOPIA and VGMIDI), the experiments show that our methods perform better in SMER task. Specifically, accuracy has been increased by 4.17 absolute point to 67.58 in EMOPIA dataset, and 1.97 absolute point to 55.85 in VGMIDI dataset. Ablation studies also show the effectiveness of multi-task methods designed in this paper.
\end{abstract}

\section{Introduction}
Emotion recognition of music has attracted lots of attention in the field of music information retrieval(MIR). For a long time, the research on music emotion recognition has been mainly carried out in the \textit{audio domain}\cite{baume2014selection,liu2018audio,panda2018novel,panda2020audio}. However, emotion recognition is less explored for music from symbolic data, such as MIDI and MusicXML formats.  Thanks to the rapid development of the symbolic music generation\cite{DBLP:conf/ismir/YangCY17,DBLP:conf/iclr/HuangVUSHSDHDE19,DBLP:conf/mm/HuangY20}, more and more research focuses on symbolic music understanding\cite{DBLP:conf/acl/ZengTWJQL21,DBLP:journals/corr/abs-2107-05223}, including symbolic music emotion recognition(SMER).

Recently, researches in SMER mainly focused on learning better representation from large-scale unlabeled music pieces via pre-training model by masked or non-masked language model borrowed from NLP and then fine-tuning the pre-trained model directly for emotion recognition on a small dataset. However, simply employing such techniques from NLP may lack the understanding of music structure which is critical to emotion classification for symbolic data\cite{DBLP:conf/acl/ZengTWJQL21}.

The existing psychology and music theory literature have revealed the relationship between music structure and emotion. Kaster \shortcite{kastner1990perception} has demonstrated that positive emotion is related to listened music in major keys, while negative emotion is related to minor keys. Similar results can be found in \cite{gerardi1995development,gregory1996development,dalla2001developmental}.  Livingstone\shortcite{livingstone2010changing} found that the loudness of music can greatly affect the expression of emotion. However, the loudness is measured in the audio domain and is still an open problem to measure it in the symbolic domain. Adli\shortcite{adli2007piano} has demonstrated that there is a linear relationship between the velocity in the symbolic domain and the loudness in the audio domain, which means that there is a connection between the velocity of music and emotion.

Recognizing the importance of musical structure for emotion recognition, we present a simple framework called MT--SMNN that incorporates the emotion recognition task with other emotion-related auxiliary tasks derived from the intrinsic structure of the music. By combining the key classification and velocity classification tasks, MT-SMNN based models can better understand emotion classification. Although MT-SMNN is a multi-task framework, we only need the manually annotated emotion label because the velocity label can be extracted directly from symbolic data, and the key label can be obtained by the well-received Krumhansl-Kessler algorithm\shortcite{krumhansl2001cognitive}, which means the proposed framework can be applied to all emotion-labeled symbolic music datasets. 

We combine the MT-SMNN framework with existing models and evaluate them in both EMOPIA and VGMIDI datasets. Results demonstrate that our proposed MT-SMNN based models achieve the new state-of-the-art on both datasets.

The chief contributions of this paper can be summarized as following aspects:
\begin{itemize}
	\item We present a novel multi-task framework called MT-SMNN, mainly focusing on emotion recognition for symbolic music. In addition to emotion recognition, a better understanding of the structure of music is also taken into account in this framework.
	\item We propose two types of auxiliary tasks for SMER. Results show that both tasks can improve the performance of SMER, especially in the valence dimension.
	\item MT-SMNN based models achieve new state-of-the-art results due to the powerful ability to learn better emotion-based knowledge from auxiliary tasks.
	\item We have reproduced most previous work for symbolic music emotion recognition on both exiting public available datasets which is helpful to building benchmarks.
\end{itemize}

\section{Related Work}\label{sec:relate_work}

We divide previous work on symbolic music emotion recognition into the following two categories.

\paragraph{Machine Learning based Methods:}Early studies used manually extracted statistical musical features and then fed them into machine learning classifiers to predict the emotion of symbolic music. Grekow \textit{et al.}\shortcite{DBLP:conf/ismis/GrekowR09} extracted 63 features from classical music in MIDI format and used \textit{k}-NN to classify the music after feature selection. Lin \textit{et al.}\shortcite{DBLP:conf/ismir/LinCY13} compared the audio, lyric , and MIDI modal of the same music, finding that MIDI modal features performed better than audio modal features in emotion recognition. Specifically, 112 types of high-level musical features were extracted from MIDI files using the JSymbolic library\cite{DBLP:conf/icmc/McKayF06}, and then SVM was employed to classify the data. Similarly, Panda \textit{et al.}\shortcite{panda2013multi} extracted 320 types of features from MIDI files using multiple tools and then classified them using SVM as well.

\paragraph{Deep Learning based Methods:} In recent years, it has become a trend to encode symbolic music into MIDI-like musical representation\cite{DBLP:journals/nca/OoreSDES20,DBLP:conf/mm/HuangY20,DBLP:conf/aaai/HsiaoLYY21} and then employ deep learning models to classify emotion.  With encoding MIDI files into MIDI-like sequences, Ferreira\cite{DBLP:conf/ismir/FerreiraW19,ferreira2020computer} used LSTM and GPT2\cite{radford2019language} for emotion classification. For simplicity, in the following, we use MIDIGPT to denote the approach proposed by \cite{ferreira2020computer}. Inspired by the great success of BERT\cite{DBLP:conf/naacl/DevlinCLT19} in NLP, Chou \textit{et al.}\shortcite{DBLP:journals/corr/abs-2107-05223} presented a large-scale pre-training model called MidiBERT-Piano, which employed CP representation\cite{DBLP:conf/aaai/HsiaoLYY21} and has shown good results in a number of fields, including symbolic music emotion recognition.

\section{Proposed Method}\label{sec:method}
\begin{figure}[htb]
	\centering
	\vspace{-1mm}
	{
		\includegraphics[width=0.5\textwidth]{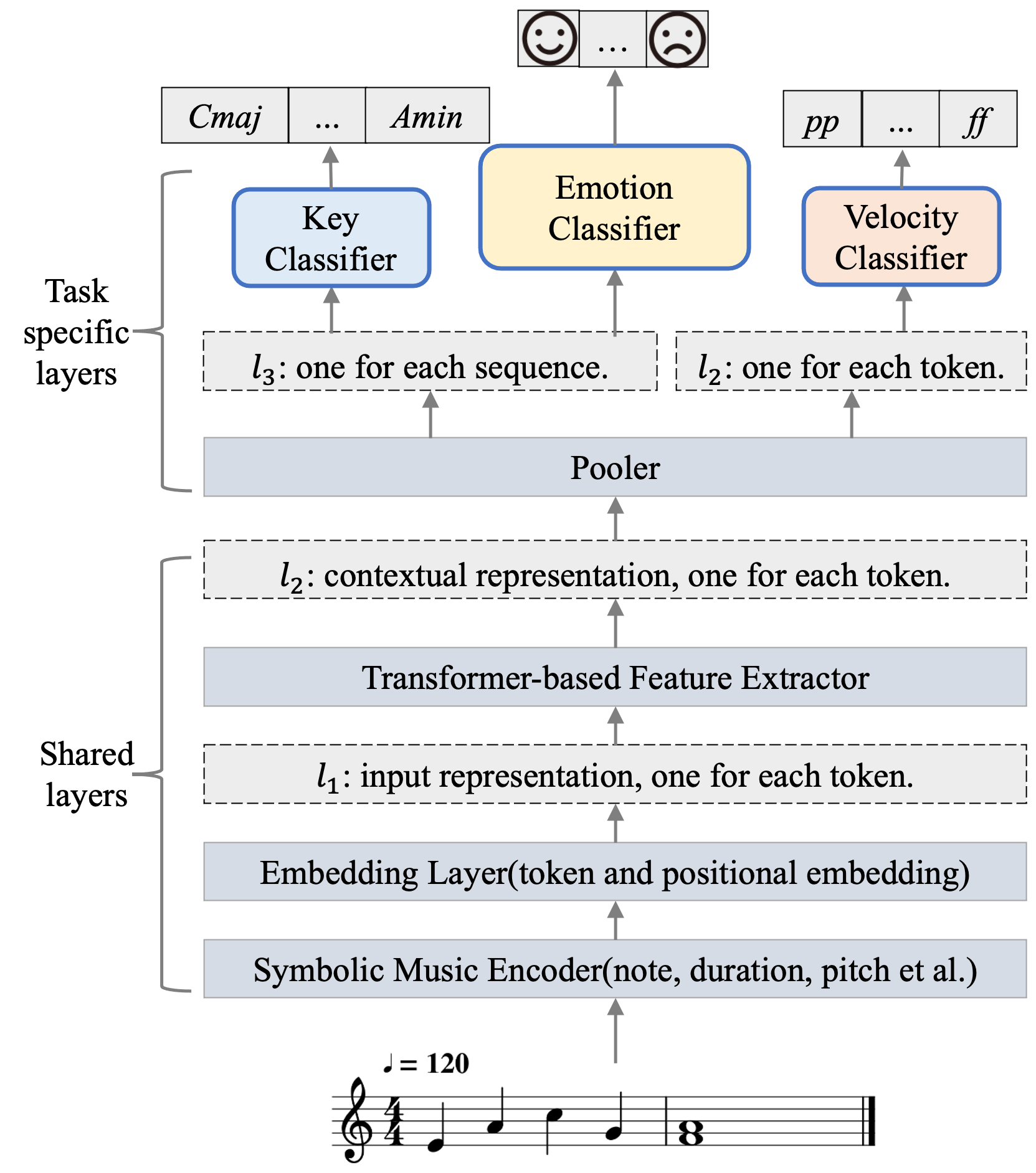}
	}
	\caption{
		The framework of proposed MT-SMNN. The text inside parentheses indicates the information processed in the corresponding module. Two \textit{sequence-level} tasks(Emotion and Key classification)and  one \textit{note-level} task(Velocity classification) are adopt in this framework.  The given music is first encoded as a sequence of musical representations and then passed to the embedding layer to get the input representation $l_1$ for each token. Next, the feature extractor captures the high-level information from each token's input representation and generates contextual representation. Then, according to whether the downstream task is sequence-level or note-level, the pooler generates the sentence representation $l_3$ or keeps origin representation $l_2$. Finally, the task-specific classifiers predict the label for each task.
	}
	\label{fig:framework}
\end{figure}

In this section, we introduce the \textbf{M}ulti-\textbf{T}ask \textbf{S}ymbolic  \textbf{M}usic \textbf{N}eural \textbf{N}etwork(MT-SMNN), a multi-task framework for symbolic music emotion recognition, as illustrated in Figure \ref{fig:framework}.  Below, we describe the structure of MT-SMNN in detail.

\subsection{Symbolic Music Encoder}\label{sec:music_encoder}

A piece of music from symbolic data, such as MIDI and MusicXML, can be encoded as a sequence of musical events, which are so called \textit{tokens} in the previous literature.  Existing method to encode symbloic music can be devided to \textit{single-word} representation and \textit{compound-word}(CP)\cite{DBLP:conf/aaai/HsiaoLYY21} representation. The MT-SMNN framework can use either single-word representation or compound-word representation. Without losing generality, we show a single-word representation method (Ferreira's\shortcite{ferreira2020computer} method) and a compound-word representation method\cite{DBLP:journals/corr/abs-2107-05223} as example in this part. We simply describe these two types of symbolic music encoding method below.

\begin{figure}[htb]
	\includegraphics[width=0.5\textwidth]{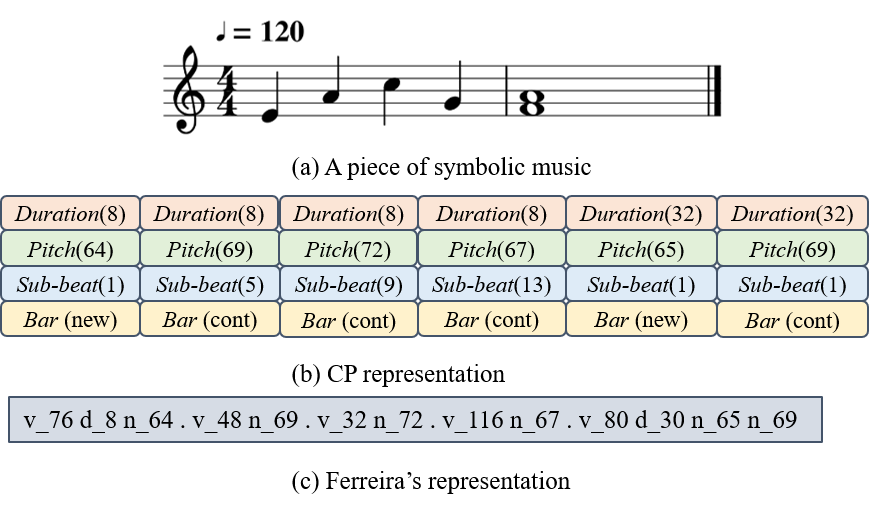}
	\caption{(a) An example of symbolic music. (b)The CP representation encodes a piece of music to a sequence of \textit{super tokens}.  Each super token consists of four \textit{sub-tokens}: \textit{Bar}, \textit{Sub-beat}, \textit{Pitch} and \textit{Duration}. The text inside parentheses means the value of the corresponding sub-token. (c) The Ferreira's method encodes a piece of music to a sequence of tokens. Token started with ``v\_``, ``d\_`` and ``n\_`` denotes the \textit{velocity}, \textit{duration} and \textit{pitch} of the corresponding note. The ``.`` token indicates the time shift event. }
	\label{fig:musical_repre}
\end{figure}

As illustrated in Figure \ref{fig:musical_repre}(b), the CP representation method encodes given piece of music to a sequence of \textit{super tokens}. Each super token consists of four \textit{sub-tokens}: \textit{Bar}, \textit{Sub-beat}, \textit{Pitch} and \textit{Duration}. The Ferreira's method, shown in Figure \ref{fig:musical_repre}(c), encodes given music as token sequence $\boldsymbol{X}=\{x^v_1, x^d_1, x^p_1, ..., x^v_S, x^d_S, x^p_S\}$, where $x^v_i$, $x^d_i$ and $x^p_i$ denotes the \textit{velocity}, \textit{duration} and \textit{pitch} for the \textit{i}-th note respectively and \textit{S} denotes the length of this sequence. 

\subsection{Embedding Layer}\label{sec:embed}
According to the symbolic music encoding method, the embedding layer of MT-SMNN can be divided into two types. The input token is directly mapped to an embedding space for the single-word representation method. Following \cite{DBLP:journals/corr/abs-2107-05223}, for the CP representation method, the embeddings of all sub-tokens inside the super token are concatenated, then fed to a linear layer to get a complete token embedding. The token embedding is added with the corresponding position embedding to capture the position information.

\subsection{Transformer-based  Feature Extractor}
The MT-SMNN employs a transformer-based model as a feature extractor. Given input representation $l_1=\{\boldsymbol{x}_0, ..., \boldsymbol{x}_S\}$,  the feature extractor generates output representation $l_2=\{\boldsymbol{h}_0, ..., \boldsymbol{h}_S\}$, where $S$ means the number of input tokens, $\boldsymbol{x}_i\in \mathbb{R}^E$,  $\boldsymbol{h}_i\in \mathbb{R}^E$  denotes the embedding and contextual representation of \textit{i}-th input token respectively, and $E$ represents the dimension of embedding  space and hidden state.

\subsection{Pooler}
Since both sequence-level and note-level tasks are employed in the MT-SMNN framework, we use a pooler to aggregate information from the entire contextual representation sequence for the sequence-level classification task. At the same time, the pooler keeps a series of contextual representations for the note-level classification task.

For simplicity, the pooler applies identical mapping for the note-level classification task. For sequence-level classification tasks, the pooler can be designed by one of the following strategies: taking the first contextual representation(like BERT\cite{DBLP:conf/naacl/DevlinCLT19}), taking the last(like MIDIGPT\cite{ferreira2020computer}), or attention-based weighting average(like MIDIBERT-Piano\cite{DBLP:journals/corr/abs-2107-05223}). In MT-SMNN, the emotion and key classification task share the same sentence representation $l_3$ because both are sequence-level tasks. 
\subsection{Task-specific Classification}
In this part, we first introduce the auxiliary tasks employed by MT-SMNN. Then, we describe more details about these classification outputs. Finally, we show the multi-task loss function used by MT-SMNN. 
\subsubsection{Auxiliary Tasks}
\paragraph{Key Classification:} This is a \textit{sequence-level} classification task.  The target is to predict the musical key for the given sequence of musical representation tokens collected from a  piece of music. There are 24 possible keys: 12 major keys and 12 minor keys\cite{thompson1997music}.

\paragraph{Velocity Classification:} This is a \textit{note-level }classification task \cite{DBLP:journals/corr/abs-2107-05223}. The target is to predict velocity for each individual note for the given sequence of notes collected from a piece of music. Following \cite{DBLP:journals/corr/abs-2107-05223}, we quantize 128 possible MIDI velocity values(0-127) into six classes: \textit{pp} (0-31), \textit{p} (32-47), \textit{mp} (48-63), \textit{mf} (64-79), \textit{f} (80-95), and \textit{ff} (96-127).

\subsubsection{Classification Ouputs}

Let $\boldsymbol{H}=\{\boldsymbol{h}_0,...,\boldsymbol{h}_S\}$ be the contextual representation($l_2$) of given piece of music, $\boldsymbol{Z}$ be the sentence representation($l_3$), where $\boldsymbol{h}_i \in \mathbb{R}^E$, $\boldsymbol{Z}\in \mathbb{R}^K$, $E$ is the dimension of hidden state , and $K$ is the dimension of sentence representation generated by the pooler. For sequence-level tasks(emotion and key classification), the probability that given a piece of music is predicted as class $c$ by a classifier with softmax can be formalized as:

\begin{equation}\label{eq:seq_pred}
	P^t(c^t|\boldsymbol{Z})=\text{softmax}(\phi^t(\boldsymbol{Z}))
\end{equation}
where $\phi$ is the mapping function of classifier, $t$ is to distinguish between different tasks.

For the note-level task(velocity classification), the probability that the \textit{i}-th note in a piece of music is predicted as class $c$ can be formalized as:

 \begin{equation}\label{eq:note_pred}
 	P^t_i(c^t|\boldsymbol{H})=\text{softmax}(\phi^t(\boldsymbol{h}_i))
 \end{equation}
 where $\phi$ is the mapping function of classifier, $i$ means the $i$-th item in coresponding sequence.

For more detail, the classifiers consist of two fully connected layers with the ReLU activation function in the middle.

\subsubsection{Multi-task loss}
For each task, we use cross-entropy loss as its objective. Let $L_1$, $L_2$, and $L_3$ be the loss of emotion recognition, key classification, and velocity classification, respectively. We employ the adaptive loss function proposed by Liebel\shortcite{liebel2018auxiliary}. The multi-task loss is formalized following:
 \begin{equation}\label{eq:multi_loss}
	L=\sum_{t} \frac{1}{2\cdot \sigma_t}L_t + \ln \left ( 1+\sigma_t^2 \right )
\end{equation}
where $L_t$ indicates the loss of task $t$, $\sigma_t$ a learnable parameter which controls the contribution of $t$-th task, and the second term is a regularizer.

\section{Experiments}
In this section, we evaluate the proposed MT-SMNN based models on EMOPIA\cite{DBLP:conf/ismir/HungCDKNY21} and VGMIDI\cite{DBLP:conf/ismir/FerreiraW19,ferreira2020computer} datasets. We first overview the datasets and processing procedure. Then, we describe the baselines and our proposed models(MT-MIDIBERT and MT-MIDIGPT). Finally, we show the results and analysis.

\begin{table}[htbp]
	\centering
	\caption{Summary of the two datasets: EMOPIA and VGMIDI.}
	\begin{tabular}{ccccc}
		\toprule
		Datasets & \#Train & \#Valid & \#Test & \#Label \\
		\midrule
		EMOPIA & 869   & 114   & 88    & 4 \\
		VGMIDI & 4,876 & 879   & 1,436 & 4 \\
		\bottomrule
	\end{tabular}%
	\label{tab:datasets}%
\end{table}%

\begin{table*}[htb] 
	\centering  
	\caption{Main result of SMER in the EMOPIA and VGMIDI dataset.}    
	\begin{tabular}{lcccc}
		\toprule    
		\multicolumn{1}{c}{\multirow{2}[3]{*}{Model}} & \multicolumn{2}{c}{EMOPIA} & \multicolumn{2}{c}{VGMIDI} \\\cmidrule{2-5}     & Accuracy(\%) & macro-F1 & Accuracy(\%) & macro-F1 \\    
		\midrule    SVM(\cite{DBLP:conf/ismir/LinCY13}) & 47.72 & 0.4763 & 45.12 & 0.3779 \\    
		SVM(\cite{panda2013multi}) & 39.77 & 0.3624 & 36.93 & 0.2146 \\    
		MIDIGPT\cite{ferreira2020computer} & 58.75$\pm$3.13 & 0.572$\pm$0.029 & 53.88$\pm$3.48 & 0.505$\pm$0.041 \\    
		MIDIBERT-Piano\cite{DBLP:journals/corr/abs-2107-05223} & 63.41$\pm$3.52 & 0.628$\pm$0.033 & 47.30$\pm$2.81 & 0.432$\pm$0.021 \\    
		\midrule    MT-MIDIGPT(proposed) & 62.50$\pm$4.45 & 0.611$\pm$0.047 & \textbf{55.85$\pm$1.97} & \textbf{0.509$\pm$0.017} \\    
		MT-MIDIBERT(proposed) & \textbf{67.58$\pm$2.39} & \textbf{0.664$\pm$0.027} & 49.81$\pm$2.52 & 0.453$\pm$0.019 \\ 
		\bottomrule
	\end{tabular} 
	\label{tab:main_result}
\end{table*}

\begin{algorithm}[tb]
	\caption{\label{algo:mt-smnn} Training a MT-SMNN-based model.}
	\begin{algorithmic}[1]
		\STATE Load model parameters $\Theta$ from the pre-trained checkpoint;
		\STATE Set the max number of epoch: $epoch_{max}$;
		\STATE Prepare dataset D;
		\FOR{$epoch$ in $1,...,epoch_{max}$}
		\FOR{$b_i$ in D}
		\STATE \textit{//$b_i$ is the \textit{i}-th mini-batch of dataset} ;
		\STATE 1. Predict \\
		\STATE \hspace{0.3cm} Predict emotion and key for music using Eq. \ref{eq:seq_pred};\\
		\STATE \hspace{0.3cm} Predict velocity for each note using Eq. \ref{eq:note_pred};\\
		\STATE 2. Compute loss \\
		\STATE \hspace{0.3cm} Compute loss  for each task using cross-entropy; \\
		\STATE \hspace{0.3cm} Compute loss $L\left( \Theta\right)$ for multi-task using Eq. \ref{eq:multi_loss};
		\STATE 4. Compute gradient: $\nabla(\Theta)$ ;
		\STATE 5. Update parameters: $\Theta = \Theta - \lambda \nabla(\Theta)$;
		\ENDFOR
		\STATE Evaluate the model in validation set;
		\STATE Check for early-stopping;
		\ENDFOR
		
	\end{algorithmic}
\end{algorithm} 

\subsection{Datasets and Preprocess}
The information of the EMOPIA and VGMIDI datasets is summarized in Table \ref{tab:datasets}. The EMOPIA dataset\footnote{https://zenodo.org/record/5257995} is a dataset of pop piano music for symbolic music emotion recognition. The clips is labeled to 4 class accoding to Russell's 4Q\cite{russell1980circumplex}. The VGMIDI dataset\footnote{https://github.com/lucasnfe/bardo-composer/tree/master/data/vgmidi} is a dataset of video game sound-tracks formatted in MIDI.  Each clip in the VGMIDI dataset is labeled as valence-arousal pair, also according to the Russell's model.

For experimental consistency, we transfer the valence-arousal pair in VGMIDI to the taxonomy of Russell's 4Q as  EMOPIA. The initial VGMIDI dataset has been split into a training set and a testing set. We divide a portion(about 15\%) of the original training set into the validation set. In this procedure, we ensure that the clips of the validation set and the training set will not come from the same song.

The MT-SMNN need two additional labels(key and velocity) besides emotion. The velocity for each note can directly derived from symbolic data. We extract the key label via the well-received Krumhansl-Kessler algorithm\shortcite{krumhansl2001cognitive}  provided by the Music21 library \cite{DBLP:conf/ismir/CuthbertA10}.

\subsection{Implementation details}\label{sec:imple_details}
For the sake of fair comparison, the vast majority of previous work mentioned in Section \ref{sec:relate_work} is reproduced. Our implementation is based on the PyTorch code open-sourced by HugginFace\cite{wolf2019huggingface}. Below, we describe the reproduced models in detail.

\subsubsection{Configuration of Machine Learning based Methods} 
In this paper, we have reproduced the machine learning based models proposed in \cite{DBLP:conf/ismir/LinCY13} and \cite{panda2013multi}. After taking the best subset of features selected in \cite{DBLP:conf/ismir/LinCY13} and  \cite{panda2013multi}, the dimension of features for Lin's method and Panda's method is 521 and 135 respectively. Both methods use the SVM classifier that works with the RBF kernel.

\subsubsection{Configuration of Deep Learning based Methods} 
\paragraph{Global Settings: } The reproduced MIDIBERT-Piano\cite{DBLP:journals/corr/abs-2107-05223}, MIDIGPT\cite{ferreira2020computer} and our proposed MT-SMNN based methods all share the following global configuration: (a) The AdamW\cite{DBLP:conf/iclr/LoshchilovH19} optimizer is adopt in this paper. The $\beta_1$, $\beta_2$ and weight decay rate is set as 0.9, 0.999 and 0.01 repectively. (b) The batch size is set as 16. (c) The learning rate is set as 3e-5 with a linear scheduler. The other trick is setting warm-up steps as 500. (d) We evaluate the models every training epoch in the validation set. The model is early stopping when the \textit{macro-F1} for emotion recognition have no improvement for $T$ consecutive epochs, where $T = \lfloor0.3*N \rfloor$, $N$ denotes the number of max training epochs. The checkpoint achieving the best metric in the validation set during the training procedure is saved and evaluated in the testing set. (e) All experiments are repeated ten times with different random seeds (from 0 to 9).

\paragraph{Specific Settings: } Following \cite{DBLP:journals/corr/abs-2107-05223}, the max sequence length of MIDIBERT-Piano is set as 512. The inside BERT model adopt BERT$_{base}$. We start fine-tuning the MIDIBERT-Piano model from the released pre-trained checkpoint\footnote{https://github.com/wazenmai/MIDI-BERT}. The max sequence length of MIDIGPT is set as 1024 and 2048 to cover the entire input sequence of tokens as much as possible when experimenting with VGMIDI and EMOPIA datasets, respectively. 
To accommodate different max sequence lengths, we pre-trained the MIDIGPT model according to \cite{ferreira2020computer}, with remaining other settings unchanged except the max sequence length. We finetune the models mentioned above at most 30 epochs in VGMIDI, and 100 epochs in EMOPIA with early-stopping discussed above.

\subsubsection{Configuration of the proposed MT-SMNN based models}
We apply the proposed MT-SMNN framework to existing deep learning based methods. For the MIDIBERT-Piano model, we extend it by combining both key classification and velocity classification with the original emotion recognition task.  However, we only incorporate the key classification with the original emotion recognition task for MIDIGPT because its representation method has already leaked the velocity information.

We coin the model that combines the proposed MT-SMNN with MIDIBERT-Piano and MIDIGPT as ``MT-MIDIBERT`` and ``MT-MIDIGPT`` respectively.

\subsection{The Training Procedure of MT-SMNN}
The training procedure of MT-SMNN is shown in Algorithm \ref{algo:mt-smnn}.  We start our training from pre-trained checkpoints, and then we finetune the MT-SMNN based model using multi-task loss. After every training epoch, we evaluate the model and check whether early-stopping.

\subsection{Comparison of state-of-the-art Methods}

We compare MT-SMNN based models with previous state-of-the-art models. The result of symbolic music emotion recognition(SMER) is shown in Table \ref{tab:main_result}.  We have reproduced all these baselines in Table \ref{tab:main_result} and described them in detail in \ref{sec:imple_details}.

Table \ref{tab:main_result} shows that the deep learning based models outperform the traditional machine learning based models. In addition, models that work with the proposed MT-SMNN framework perform better than the counterpart for single-task and achieve new state-of-the-art results. Specifically, Compared with the MIDIBERT-Piano model, the proposed MT-MIDIBERT model pushes the accuracy to 67.58\% and 49.8\%, which amounts 4.2\% and 2.5\% absolution improvement on the EMOPIA and VGMIDI dataset, respectively. The proposed MT-MIDIGPT model also improves the accuracy by 3.8\% and 2.0\% to 62.50\% and 55.85\% for these two datasets, respectively. 

Since the MT-SMNN based models have no difference except for multiple classifiers, which have a minimal amount of parameters, are employed for different tasks compared with its single-task counterpart, the improvement of the above results is attributed to our proposed MT-SMNN framework.

\subsection{Ablation Studies}
In this section, we conduct experiments on the EMOPIA dataset to study auxiliary tasks' impact. The results are summarized in Table \ref{tab:aux_tasks}. 

\begin{table}[H]
	\centering
	\caption{The contribution of different tasks on EMOPIA dataset using MIDIBERT backbone. Results are evaluated by the accurary of music emotion recognition task.}
	\begin{tabular}{ccc}
		\toprule
		Key Classification & Velocity Classificaiton & Accuracy \\
		\midrule
		\xmark    & \xmark     & 63.41$\pm$3.52 \\
		\cmark     & \xmark     & 67.03$\pm$2.54 \\
		\xmark     & \cmark     & 64.73$\pm$5.47 \\
		\cmark     & \cmark  & \textbf{67.58$\pm$2.39} \\
		\bottomrule
	\end{tabular}%
	\label{tab:aux_tasks}%
\end{table}%

Table \ref{tab:aux_tasks} shows that both key and velocity classification auxiliary tasks effectively affect emotion recognition. Moreover, the model taken in both auxiliary tasks outperforms models only taken in a single. The accuracy is increased by 3.6\% and 1.3\% to 67.03\% and 64.73\% after combing the SMER task with the key and velocity classification task, respectively, which means that the key classification task is a more critical auxiliary task than the velocity classification task.

\begin{figure}[htb]
	\includegraphics[width=0.5\textwidth]{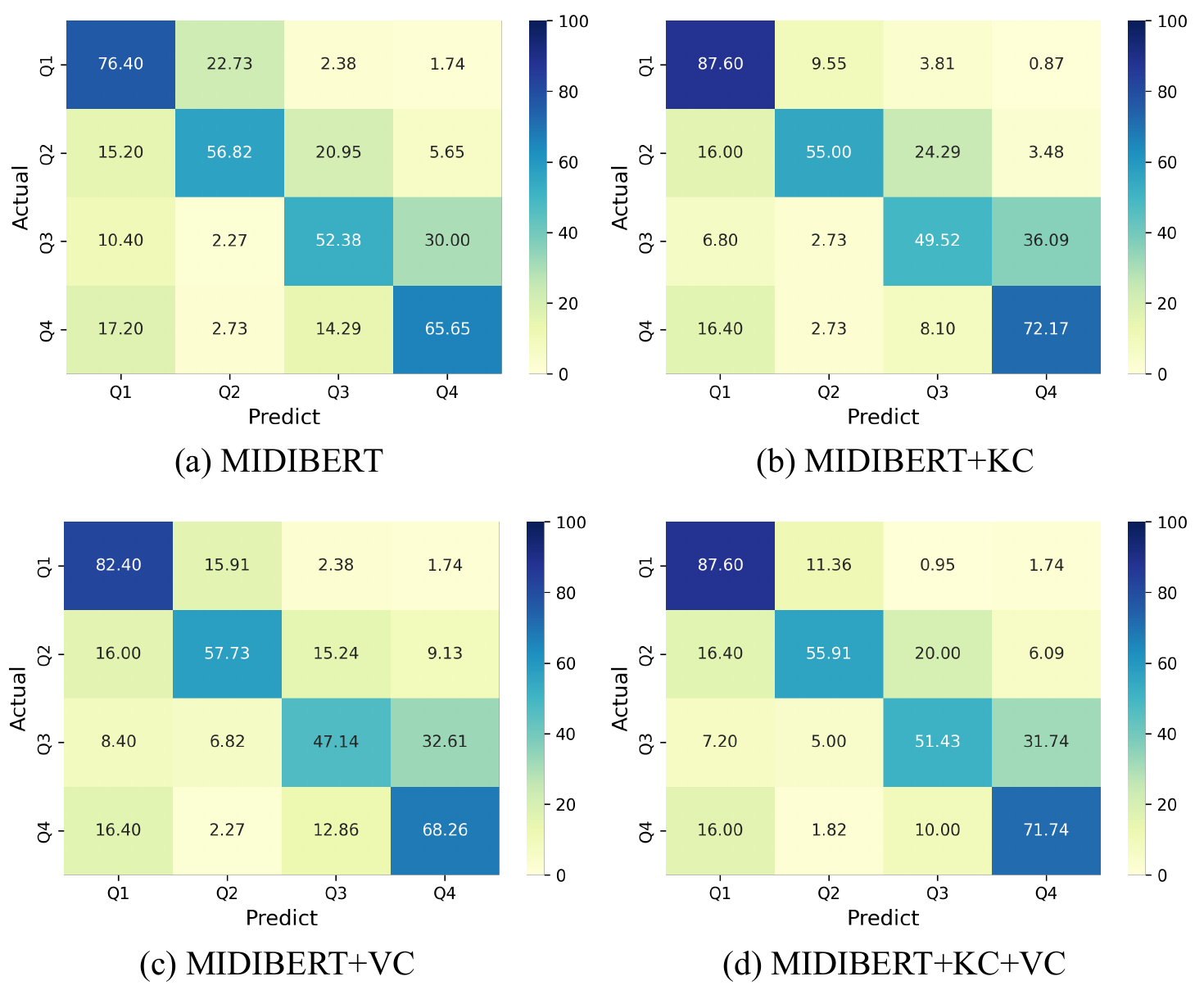}
	\caption{The result of baseline model(MIDIBERT-Piano, abbreviated to MIDIBERT) is shown in (a). The performance of models incorporate with addtional Key Classifier(KC) or Velocity Classifier(VC) is shown in (b) (c) and (d). }
	\label{fig:cm}
\end{figure}

We also plot the confusion matrices of these experiments, as shown in Figure \ref{fig:cm}. In this figure, \textit{Q1}, \textit{Q2}, \textit{Q3} and \textit{Q4} denotes \textit{HVHA}(high valence high arousal), \textit{LVHA}(low valence high arousal), \textit{LVLA}(low valence low arousal) and \textit{HVLA}(high valence low arousal) respectively which also so-called \textit{Happy}, \textit{Angry}, \textit{Sad} and \textit{Calm}  in some literatures. Compared Figure \ref{fig:cm}(b) with Figure \ref{fig:cm}(a), we have found that the key classification task can greatly improve the performance of emotion recognition in the class of Q1 and Q4. Similar results can be found in Figure \ref{fig:cm}(c) and Figure\ref{fig:cm}(d). Since Q1 and Q4 are both in the high valence region, we finally conclude that our proposed MT-SMNN framework can improve the performance of music recognition, especially in the valence dimension.

\section{Conclusion}
In this paper, we present MT-SMNN, a multi-task framework that mainly
focus on emotion recognition for symbolic music. The MT-SMNN framework combines emotion recognition with key classification and velocity classification tasks and conducts a multi-task training procedure in a single dataset. MT-SMNN based models obtain new state-of-the-art results in both EMOPIA and VGMIDI datasets. Further analysis also verifies the effectiveness of both auxiliary tasks. 

We would like to apply the MT-SMNN framework to other areas for future work. For example, the MT-SMNN based models can be employed to build a metric for evaluating the performance of emotion conditioned symbolic music generation models.

\section*{Acknowledgements}

The authors would like to thank Yi-Hsuan Yang for the open-source code of MIDIBERT-Piano\cite{DBLP:journals/corr/abs-2107-05223} and the dataset of EMOPIA\cite{DBLP:conf/ismir/HungCDKNY21}, Lucas N. Ferreira for making the dataset of VGMIDI\cite{ferreira2020computer} public available.

{
	\small
	\bibliographystyle{named}
	\bibliography{ijcai22}
}
\end{document}